\documentclass{iau}
\usepackage{graphicx}

\title[Neutron star magnetic field of two Z sources] %% give here short title %%
{Determining the neutron star surface magnetic field strength of two Z sources}

\author[Guoqiang Ding, Chunping Huang \& Yanan Wang]   %% give here short author list %%
{Guoqiang Ding$^1$, Chunping Huang$^{1,2}$, 
\and Yanan Wang$^{1,2}$}

\affiliation{$^1$Xinjiang Astronomical Observatory, Chinese Academy of Sciences, \\  150, 
Science 1-Street, Urumqi, Xinjiang 830011, 
China \\ email: {\tt dinggq@gmail.com } \\[\affilskip]
$^2$University of Chinese Academy of Sciences, \\ China}

\pubyear{2012}
\volume{290}  %% insert here IAU Symposium No.
\jname{Feeding compact objects: Accretion on all scales}
\editors{C.M. Zhang, T. Belloni, M. M\'endez \& S.N. Zhang, eds.}

\begin{document}

\maketitle

\begin{abstract}
From the extreme position of disk motion, we infer the neutron star (NS) surface 
magnetic field strength ($B_0$) of Z-source GX 17+2 and Cyg X-2. The inferred $B_0$ 
of GX~17+2 and Cyg~X-2 are $\sim$(1--5)$\times10^8\ {\rm G}$ 
and $\sim$(1--3)$\times10^8\ {\rm G}$, respectively, which are not inferior to that of 
millisecond X-ray pulsars or atoll sources. It is likely that the NS magnetic axis of 
Z sources is parallel to the axis of rotation, which could result in the lack of 
pulsations in these sources.

\end{abstract}

\firstsection % if your document starts with a section,
\section{Introduction}

The neutron star (NS) surface magnetic field strength ($B_0$) could be a critical 
parameter responsible for the behaviors of NS X-ray binaries (NSXB). However, it is 
difficult to measure $B_0$ directly. Nevertheless, it is feasible to estimate $B_0$ 
from some observable phenomena, such as quasi-periodic oscillations (QPOs) (e.g. 
\cite[Fock 1996]{Fock1996}) or ``propeller'' effects (e.g. \cite[Zhang et al. 
1998]{Zhang1998}), or inferred from theoretical models (\cite[Zhang 
\& Kojima 2006]{Zhang2006}). From previous studies, it is believed that $B_0$ is 
larger in Z sources than in atoll sources (ref. \cite[Fock 1996]{Fock1996}, 
\cite[Zhang et al. 1998]{Zhang1998}, \cite[Chen et al. 2006]{Chen2006}, \cite[Ding 
et al. 2011]{Ding2011}) or millisecond X-ray pulsars (ref. \cite [Cackett et al. 
2009]{Cackett2009}, \cite[Di Salvo \& Burderi 2003]{DiSalvo2003}). Unlike in 
millisecond X-ray pulsars or atoll sources, pulsations have never been observed in 
Z sources, why? 

\section{Method and Data Analysis}

As discussed by \cite{Zhang1998}, during the regular accretion state the inner disk 
radius ($R_{\rm in}$) of an NSXB, which is equal to the magnetospheric radius ($R_{\rm m}$) (\cite[Lamb et al. 1973]{Lamb1973}), will vary between the radius of the 
innermost stable circular orbit (ISCO) and the corotation radius ($R_{\rm c}$). 
Therefore, assuming $R_{\rm in}=R_{\rm c}$, we will get the upper limit of $B_0$, 
because more magnetic pressure is needed to push the accretion disk further, having 
\begin{equation}
B_{\rm 0,\ max} = 1.5\times 10^{12}\left(\frac{\dot{M}_{\rm m}}{\dot{M}_{\rm Edd}}
\right)^{1/2}\left(\frac{M_{\rm ns}}{1.4\mbox{ }M_{\odot}}\right)^{5/6} \left(\frac{R_{\rm 
ns}}{10^6\mbox{ }{\rm cm}}\right)^{-3}f^{-7/6}{\xi_{1}}^{-7/4}\mbox{ }{\rm G},
\end{equation}
where $f$ is the NS spin frequency and $\xi_{1}$ is a coefficient in the range 
of 0.87--0.95 (\cite[Wang 1995]{Wang1995}). Similarly, 
letting $R_{\rm in}=R_{\rm ISCO}$, we will obtain the lower limit of $B_0$, because 
less magnetic pressure is needed to balance the gas pressure from the disk, getting       
\begin{equation}
B_{\rm 0,\ min} = 2.8\times 10^{8}\left(\frac{\dot{M}_{\rm m}}{\dot{M}_{\rm Edd}}
\right)^{1/2}\left(\frac{M_{\rm ns}}{1.4\mbox{ }M_{\odot}}\right)^{2} \left(\frac{R_{\rm 
ns}}{10^6\mbox{ }{\rm cm}}\right)^{-3}{\xi_{1}}^{-7/4}\mbox{ }{\rm G}.
\end{equation}
With software HEASOFT 6.11 and FTOOLS V6.11, we use the {\it RXTE} observations made 
during Oct. 3--12, 1999, a total of $\sim$297.6 ks for GX 17+2 and those during Jul. 
2--7, 1998, a total of $\sim$42 ks for Cyg X-2 to perform our analysis. We divide the 
track on the hardness-intensity diagrams (HIDs) into some regions and then produce the 
spectrum of each region. XSPEC version 12.7 is used to fit the spectra with spectral 
models. From the spectral fitting parameters, we calculate mass accretion rates. Then, 
with the known NS mass and radius, making use of the above equations, we 
calculate the limits of $B_0$.

\section{Result and Discussion}

With mass accretion rates and $f_{\rm max}=584\ {\rm Hz}$ for GX 17+2 
and $f_{\rm max}=658\ {\rm Hz}$ for Cyg X-2 (\cite[Yin et al. 2007]{Yin2007}), we 
infer the limits of $B_0$ and get $(1\leq B_0\leq 5)\times10^8\ {\rm G}$ 
and $(1\leq B_0\leq 3)\times10^8\ {\rm G}$ for GX~17+2 and Cyg~X-2, respectively, 
which are higher than the reported $B_{\rm 0}\sim(0.3-1)\times10^8$~G of atoll 
source Aql~X-1 (\cite[Zhang et al. 1998]{Zhang1998}), compatible 
with $B_{\rm 0}\sim(1-5)\times10^8$~G of accreting millisecond X-ray pulsar 
SAX~J1808.4--3658 (\cite[Di Salvo \& Burderi 2003]{DiSalvo2003}), but lower 
than $B_{\rm 0}\sim(1-3)\times10^9$~G of the first transient Z source 
XTE J1701-462 (\cite[Ding et al. 2011]{Ding2011}) 
or $B_{\rm 0}\sim(1-8)\times10^9$~G of Cir X-1 (\cite[Ding et al. 2006]{Ding2006}). 

Since the $B_0$ of Z sources is not inferior to that of millisecond X-ray pulsars or 
atoll source, pulsations should have been observed in the former, as in the latter. 
However, pulsations have not been detected in Z sources, why? \cite[Pringle \& Rees 
(1972)]{Pringle1972} proposed that the NS pulsation emission would depend on the shape 
of emission cone, the orientations of the magnetic and rotation axes, and the line of 
the sight, and, furthermore, \cite[Lamb et al. (1973)]{Lamb1973} suggested that the 
detected pulsation could be resulted from the condition that the NS magnetic axis does 
not coincide in direction with the axis of rotation. Therefore, it is likely that the 
NS magnetic axis of Z sources is parallel to the axis of rotation, or the orientation 
of the rotation axis is in agreement with the line of the sight, any of which could 
result in lack of pulsations in these sources. 

\section{Acknowledgements}

This work is supported by the National Basic Research Program of China (973 Program 
2009CB824800) and the Natural Science Foundation of China under grant no. 11143013.

\end{document}